\documentclass[11pt]{amsart}
\usepackage{geometry}                
\usepackage{graphicx}
\usepackage{amssymb}
\usepackage{epstopdf}
\newcommand{\beq}{\begin{equation}}
\newcommand{\eeq}{\end{equation}}
\newcommand{\beqa}{\begin{eqnarray}}
\newcommand{\eeqa}{\end{eqnarray}}

\title[Analytical solution for a class of inverse problems]{Beyond inverse Ising model: structure of the analytical solution for a class of inverse problems}
\author{Iacopo Mastromatteo}
\address{Iacopo Mastromatteo\\ {\it SISSA, Via Bonomea 265, 34136 Trieste, Italy }}

\thanks{I acknowledge M.Marsili, G.Gori and S.Cocco for very useful discussions}

\begin{document}
\maketitle
\begin{abstract}
I consider the problem of deriving couplings of a statistical model from measured correlations, a task which generalizes the well-known inverse Ising problem. After reminding that such problem can be mapped on the one of expressing the entropy of a system as a function of its corresponding observables, I show the conditions under which this can be done without resorting to iterative algorithms. I find that inverse problems are \emph{local} (the inverse Fisher information is sparse) whenever the corresponding models have a factorized form, and the entropy can be split in a sum of small cluster contributions. I illustrate these ideas through two examples (the Ising model on a tree and the one-dimensional periodic chain with arbitrary order interaction) and support the results with numerical simulations. The extension of these methods to more general scenarios is finally discussed.
\end{abstract}

\section{Introduction}
The estimation of the direct interactions among the microscopic constituents of an extended system is a problem that has recently received considerable interest from the literature of several communities (biology \cite{Socolich:2005vy,Weigt:2009on}, genetics \cite{Braunstein:2008sa,Bailly-Bechet:2010kx}, neuroscience \cite{Schneidman:2006vg,Shlens:2006uq,Cocco:2009mb}, economy \cite{Lillo:2008ht,Moro:2009wy,Lachapelle:2010kv}). This is mainly due to the availability of large datasets across various fields, which have introduced the possibility to directly fit from data the interaction structure (e.g., the wiring pattern of neurons, the regulatory network of genes, the cross-influence of traders) of a system which is operating according to an unknown, or a partially known, mechanism.
Despite this abundance of data, the problem of estimating the most appropriate model in order to describe a system is far from being solved:
even if the maximum entropy principle (MEP) is invoked in order deal with the issue of model selection  \cite{Jaynes:1957uq,Jaynes:1957fk}, one is typically left with a difficult inference problem. In particular fitting a model under the MEP requires matching a set of observables estimated from data with the ones predicted by the model (the so-called \emph{inverse problem}), a task which is known to be generically hard \cite{Wainwright:2008kx}.
While the \emph{Boltzmann learning} algorithm can be used to obtain accurate solutions for the inverse problem only by investing a large amount of computational power \cite{Ackley:1985hc}, mean-field techniques \cite{Kappen:1998dt,Tanaka:1998tg,Sessak:2009lf,Roudi:2009qm,Roudi:2011rr,Ricci-Tersenghi:2012uq}, message-passing methods \cite{Mezard:2009ul,Higuchi:2009lq,Aurell:2010vn,Marinari:2010vn} and pseudo-likelihood approximations \cite{Wainwright:2007zr,Aurell:2012ys} have proved to be very effective in order to fit approximately the interactions of the system in a short amount of time. \\ \\
In this work I show that for a specific class of models the inverse problem can be solved exactly and efficiently. In particular, if the probability distribution of the system has a factorized form, the entropy of the system can be written as a sum of cluster contributions which are easy to manipulate analytically. This is consistent with the results of  \cite{Cocco:2011vn,Cocco:2012uq}, who show that efficient, fast inference procedures can be constructed if the entropy of the system can be approximated by the sum of a suitable set of small cluster contributions. For these systems one can see that the robustness of the inverse problem is linked with the factorization property of the probability distribution (equivalently, the additivity of the entropy). This is a consequence of a general fact: inverse problems are typically more stable with respect to the direct ones because the response  of couplings due to a change in the observables is much more localized than the response of the observables due to a shift of the couplings. \\ \\
In the following section I introduce a very generic setup for the inverse problem, in order to stress the generality of the results. In section \ref{sec:TwoModels} its exact solution is presented for two specific cases. In \ref{sec:Appl} the results are scrutinized against numerical simulations involving synthetic data, and in \ref{sec:General} the method introduced is extended to more general cases. In section \ref{sec:Concl} some conclusions are drawn.

\section{The inverse problem\label{sec:Inv}}
\begin{figure}[h]
   \centering
   \includegraphics[width=3in]{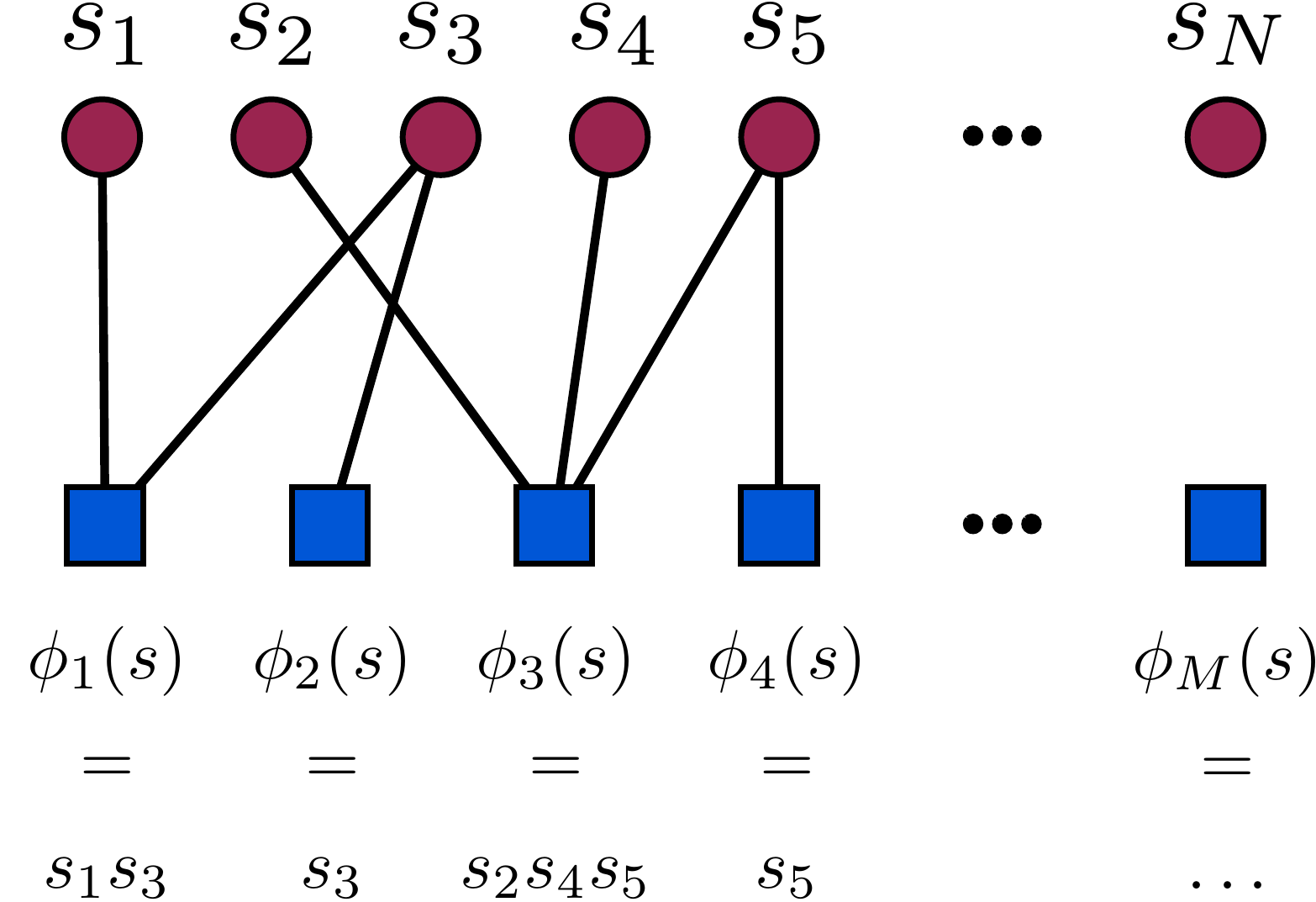} 
   \caption{Example of a typical factor graph associated with the probability density \eqref{eq:PDens}.}
   \label{fig:System}
\end{figure}

I consider the problem of estimating a set of $M$ couplings $g= ( g_1,\dots,g_M )$ describing a probability density for a set of $N$ binary spins $s=(s_1,\dots, s_N ) \in \{ -1,1\}^N$ of the form
\beq
p(s) = \frac{1}{Z(g)} \exp \left( \sum_{\mu=1}^M g_\mu \phi_\mu (s) \right) \; , \label{eq:PDens}
\eeq 
where $\phi = (\phi_1, \dots, \phi_M)$ is a set of known functions of $s$ (usually referred as \emph{potentials}, or \emph{sufficient statistics}), and the \emph{partition function} $Z(g)$ enforces the normalization of $p(s)$. In the following discussion we will assume all of the $\phi_\mu (s)$ to be monomials, i.e., to be functions of the form
\beq
\phi_\mu (s) = \prod_{i \in \Gamma_\mu} s_i \; ,
\eeq
where $\Gamma_\mu$ is a given subset (cluster) of the \emph{vertex} set $V=\{ 1,\dots N\}$  associated with the operator $\mu$.
Within this general formulation, the Ising (or \emph{graphical}) model can for example be constructed by using single-body clusters $\{ i \}_{i=1}^N$ and two-body clusters $\{ i,j\}_{i<j=1}^N$, together with their associated couplings usually denoted with $\{ h_i \}_{i=1}^N$ and $\{J_{ij}\}_{i<j=1}^N$. Figure \ref{fig:System} shows factor graph associated with the probability density \eqref{eq:PDens} in a generic case.
In order to estimate such model, a set of $T$ i.i.d. observations $\hat s = \{ s^{(t)} \}_{t=1}^T$of the system is provided, so that the log-likelihood function for the model can be written as
\beq
\log L(\hat s | g) = T \left( - \log Z(g) + \sum_{\mu=1}^M g_\mu \bar \phi_\mu \right) \,
\eeq
and depends on the data through the empirical averages $\bar \phi_\mu = \frac{1}{T} \sum_{t=1}^T \phi_\mu (s^{(t)})$. Maximizing such function in order to find the coupling set $g^\star$ best describing the data $\hat s$ in absence of prior leads to the set of $M$ conditions
\beq
\langle \phi_\mu \rangle_g = \bar \phi_\mu \; ,
\eeq 
where the quantities $\langle \phi_\mu \rangle_g = \partial \log Z(g) / \partial g_\mu$ describe the averages of the functions $\phi_\mu(s)$ under the model parametrized by $g$. Thus, in order to find the best model in order to describe data $\hat s$ one has to match empirical averages  $\bar \phi_\mu$ with ensemble averages $\langle \phi_\mu \rangle_{g^\star}$. An equivalent characterization of the problem of finding the optimal $g^\star$ is provided by computing the Legendre transform
\beq
-S(\bar \phi) = \max_g \left( \sum_{\mu=1}^M g_\mu \bar \phi_\mu -\log Z ( g) \right) \; ,
\eeq
which can be identified as the Shannon entropy of the optimal model expressed as a function of the empirical averages $\bar \phi$. Such quantity is the cumulant generating function for the maximum likelihood estimator of the couplings $g^\star = g^\star(\bar \phi)$. From this perspective, the entropy $S(\bar \phi)$ plays a role similar to the one of the free-energy $F(g) = -\log Z(g)$ in a statistical mechanics problem (i.e., estimating the ensemble averages $\langle \phi_\mu \rangle_g$ given the couplings $g$). In fact, $F(g)$ satisfies the relations
\beqa
\langle \phi_\mu \rangle_g &=& - \frac{\partial F(g)}{\partial g_\mu} \\
\chi_{\mu,\nu}(g) &=& - \frac{\partial^2 F(g)}{\partial g_\mu \partial g_\nu} \,
\eeqa
being $\chi_{\mu,\nu}(g) $ the covariance matrix $ \langle \phi_\mu \phi_\nu \rangle_g - \langle \phi_\mu \rangle_g \langle \phi_\nu \rangle_g $. Such matrix $\chi_{\mu,\nu}$ is a central object in the field of information theory, where it is customarily referred as \emph{Fisher information} \cite{Cover:1991fk,Mezard:2009ko}. For $S(\bar \phi)$ it holds
\beqa
g^\star_\mu (\bar \phi)&=& - \frac{\partial S(\bar \phi)}{\partial \bar \phi_\mu} \label{eq:GenFuncEst}\\
\chi_{\mu,\nu}^{-1}(\bar \phi) &=& - \frac{\partial^2 S(\bar \phi)}{\partial \bar \phi_\mu \partial \bar \phi_\nu} \label{eq:InvFish}\, \; .
\eeqa
Just as the quantity $\chi_{\mu,\nu}(g)$ can be used to express how fast in $T$ the empirical averages $\bar \phi$ converge to the ensemble values $\langle \phi \rangle$, its matrix inverse expresses the  rate of convergence of the inferred couplings $g^\star(\bar \phi)$ to the asymptotic values $g^\star(\langle \phi \rangle)$. In particular the covariance matrix for the estimator $g^\star(\bar \phi)$ is given by 
\beq
\bigg< (g_\mu^\star(\bar \phi)-g_\mu^\star(\langle \phi \rangle)) (g_\nu^\star(\bar \phi)-g_\nu^\star(\langle \phi \rangle)) \bigg>_T - \bigg< g_\mu^\star(\bar \phi) - 
g^\star_\mu(\langle \phi \rangle) \bigg>_T \bigg< g_\nu^\star(\bar \phi) - g^\star_\nu(\langle \phi \rangle) \bigg>_T  = \frac{\chi^{-1}_{\mu,\nu}}{T} \, \; ,
\eeq
where $\langle \cdot \rangle_T = \sum_{\hat s} \prod_{t=1}^T p(s^{(t)})$. Hence the matrix $\chi^{-1}(\bar \phi)$ expresses the stability of the inferred couplings with respect to the statistical error due to finite sampling. \\ \\
I will show in the following section how equations \eqref{eq:GenFuncEst} and \eqref{eq:InvFish} can be used to determine analytically $g^\star(\bar \phi)$. In particular, if it is possible to compactly express the entropy of a model as a function of the empirical averages $\bar \phi$, the maximum likelihood estimator $g^\star(\bar \phi)$ can be found easily.

\section{Two factorizable models \label{sec:TwoModels}}
In this section I present two models for which the inverse problem can be exactly solved. The solution strategy can be straightforwardly extended to more general cases. \\ \\
In order to discuss this approach to the inverse problem, it is nevertheless necessary to introduce the notion of \emph{marginal} $p_\Gamma$, which given a set $\Gamma \subseteq V$ is defined as
\beq
p_\Gamma(s_\Gamma) = \sum_{s_i | i \not\in \Gamma} p(s) 
\eeq
and allows to define the \emph{cluster entropy} $S_\Gamma$ as
\beq
S_\Gamma(p_\Gamma) = - \sum_{s_\Gamma} p_\Gamma(s_\Gamma) \log p_\Gamma (s_\Gamma) \; .
\eeq
It is easy to prove that marginals and cluster entropies can be exactly expressed as functions of the averages of all the possible monomials contained in the cluster $\Gamma$. More precisely, given the set of $2^{|\Gamma|}$ monomials $\{ \hat \phi_{\Gamma^\prime} \}_{\Gamma^\prime \subseteq \Gamma}$ associated with the clusters $\Gamma^\prime \subseteq \Gamma$, it is easy to show that
\beq
p_\Gamma(s_\Gamma) = \frac{1}{2^{|\Gamma|}} \sum_{\Gamma^\prime \subseteq \Gamma} \hat \phi_{\Gamma^\prime} (s_{\Gamma^\prime}) \langle \hat \phi_{\Gamma^\prime} \rangle_g \; . \label{eq:Marg}
\eeq
This formula characterizes the marginals $p_\Gamma(s_\Gamma)$ as \emph{local} objects, as they are uniquely determined by the averages of potentials contained inside $\Gamma$.

\subsection{Tree-like graphs}
Consider a model of the form
\beq
p(s) = \frac{1}{Z(h,J)} \exp \left( \sum_{i \in V} h_i s_i + \sum_{(i,j) \in E} J_{ij} s_i s_j \right) \, , \label{eq:IsingMod}
\eeq
where $E = \{ (i,j) \in V \times V \; |\; i < j \}$ is a given set of \emph{edges}. Suppose additionally that the graph associated with the edge set $E$ is acyclic, so that the model \eqref{eq:IsingMod} corresponds to an Ising model on a tree (more precisely, a forest). In that case, it can be shown by induction (see \cite{Mezard:2009ko,Wainwright:2008kx}) that the total probability density $p(s)$ can be factorized according to
\beq
p(s) = \prod_{(i,j) \in E} p_{\{i,j\}}(s_i,s_j) \prod_{i\in V} \left[ p_{\{i\}} (s_i) \right]^{1- |\partial i|} \, , \label{eq:TreeFact}
\eeq
where $\partial i = \{ (j,k) \in E \; | \;  i \in (j,k) \}$.
Then, the entropy of the model is additive and can be written as
\beq
S(p) = \sum_{(i,j) \in E} S_{\{i,j\}}(m_i,m_j,c_{ij}) + \sum_{i \in V} (1-|\partial i|)S_{\{ i\}}(m_i) \; ,
\eeq
where one has defined the \emph{magnetizations} $m_i = \frac{1}{T} \sum_{t=1}^T s_i^{(t)}$ and the \emph{correlations} $c_{ij} = \frac{1}{T} \sum_{t=1}^T s_i^{(t)} s_j^{(t)}$.
After the use of formula \eqref{eq:Marg}, one finds by differentiation that the solution of the inverse Ising model on an acyclic graph is given by
\beqa
h_i^\star &=& \frac{1}{4} \sum_{j \in \partial i}  \sum_{s_i,s_j}  s_i \log \Bigg[ \frac{1}{4}(1 + m_i s_i + m_j s_j + c_{ij} s_i s_j) \Bigg]  \nonumber \\
&+& \frac{1}{2} (1-|\partial i |) \sum_{s_i}  s_i \log \Bigg[ \frac{1}{2}(1 + m_i s_i ) \Bigg]  \label{eq:SolTree} \\
J_{ij}^\star &=& \frac{1}{4} \sum_{s_i,s_j}  s_i s_j \log \Bigg[ \frac{1}{4}(1 + m_i s_i + m_j s_j + c_{ij} s_i s_j) \Bigg]  \; , \nonumber
\eeqa
and the stability of the solution is determined by
\beqa
\chi^{-1}_{\{i,j\},\{k,l\}} &=& \frac{1}{16} \sum_{s_i,s_j} \frac{ \delta_{i,k} \delta_{j,l} + \delta_{i,l} \delta_{j,k}}
{\bar p_{\{i,j\}}(s_i,s_j)}  \nonumber \\
\chi^{-1}_{\{i,j\},\{k\}} &=&  \frac{1}{16} \sum_{s_i,s_j} \frac{\delta_{i,k} s_j + \delta_{j,k} s_i}{\bar p_{\{i,j\}}(s_i,s_j)} \label{eq:FisherInfoPairTree} \\
\chi^{-1}_{\{i\},\{j\}} &=& \frac{1}{16} \sum_{k \in \partial i} \sum_{s_i s_k} \frac{ \delta_{i,j} + s_i s_k \delta_{k,j} }{\bar p_{\{i,k\}}(s_i,s_k)} +
\frac{1}{4} (1-|\partial i|) \sum_{s_i} \frac{\delta_{i,j}}{\bar p_{\{i\}}(s_i)} \nonumber \; .
\eeqa
The quantities
\beqa
\bar p_{\{i\}}(s_i) &=& \frac{1}{2} \left( 1 + m_i s_i \right) \\
\bar p_{\{i,j\}}(s_i,s_j) &=& \frac{1}{4} \left( 1 + m_i s_i + m_j s_j + c_{ij} s_i s_j\right)
\eeqa
appearing in the previous expression describe the empirical frequencies associated with the clusters $\{ i \}$ and $\{ i,j\}$. \\ \\
Notice that the problem of finding any of the mean magnetizations $\langle s_i \rangle_{(h,J)}$ or a correlations $\langle s_i s_j \rangle_{(h,J)}$ usually requires the use of iterative algorithms such as belief propagation, which converge in a number of steps linear with the number of vertices $|V|$ \cite{Mezard:2009ko}. For the inverse problem no iterative algorithm is required, as the solution can be found by simply evaluating formula \eqref{eq:SolTree}. This is a consequence of the fact that -- as  observed in \cite{Cocco:2011vn,Cocco:2012uq} -- inverse problems tend to be more local than their direct counterparts. Equivalently, due to sparsity of $\chi^{-1}$, by shifting an empirical average (either $m_i$ or $J_{ij}$) just neighboring couplings are changed, whereas for the direct problem the change of a coupling (say, $h_i$), has an effect on a finite number of vertices (being the size of the perturbation roughly determined by the correlation length of the system).
These results are consistent with the ones of \cite{Ricci-Tersenghi:2012uq,Welling:2003kx}, where the solution of the inverse Ising problem under the Bethe approximation is derived. In particular, equation \eqref{eq:SolTree} can be recovered by specializing the results in \cite{Ricci-Tersenghi:2012uq,Welling:2003kx} to the case of a tree, for which the Bethe approximation is exact.
\\ \\
This solution requires indeed the knowledge of the full topology specified by $E$, so that this approach cannot be used to determine the graph structure from data. Rather, it allows to exactly recover the coupling strength given an acyclic set of edges $E$.

\subsection{One-dimensional models}
Within this approach it is also possible to find the exact solution of the inverse problem for a one-dimensional chain with interactions of arbitrary order.
Even if a heuristic solution for this model was first proposed in \cite{Gori:2011ly}, its detailed derivation is provided in the following.
The model is defined by a probability distribution $p(s)$ on a set of  binary spins $s \in \{ -1,1\}^N$ specified by a family of potentials acting on the first $R$ spins $\phi (s_1,\dots , s_R) = (\phi_1(s_1,\dots, s_R), \dots, \phi_{M}(s_1,\dots, s_R))$ subject to the periodic boundary conditions $s_i = s_{i+N}$. By using \emph{translation operators} $T=\{T_n\}_{n=0}^{N/\rho-1}$, defined through their action on the $\phi$
\beq
T_n \phi_\mu (s_1, \dots s_R) = \phi_\mu(s_{1+n \rho}, \dots , s_{R+n \rho}) \; , \label{eq:OneDimChain}
\eeq
one can define a \emph{one-dimensional chain}  as the probability distribution
\beq
p(s) = \frac{1}{Z(g)} \exp \left( \sum_{\mu=1}^{M} g_\mu \sum_{n=0}^{N/\rho-1} T_n \phi_\mu (s) \right) \; . \label{eq:1DimChain}
\eeq
Suppose that the family $\phi=(\phi_1,\dots,\phi_M)$ contains all and only the $2^{R}(1-2^{-\rho})$ monomials describing the unit cell (i.e., any potential of range smaller than $R$ along the chain can be uniquely obtained by translating one of the $\phi_\mu$ by mean of a $T_n$ operator). Then one can exploit the factorization
\beq
p(s) = \prod_n \frac{p_{\Gamma_n}(s_{\Gamma_n})}{p_{\gamma_n}(s_{\gamma_n})} \label{eq:ChainFact} \; ,
\eeq
where one has defined the sets  $\Gamma_n = \{ n\rho+1, \dots,n\rho+R\}$ and $\gamma_n=\{ (n+1)\rho+1,\dots, n\rho+R\}$.
This leads to the additivity formula for the entropy
\beq
S(p) = \sum_n \big[ S_{\Gamma_n}(p_{\Gamma_n}) -S_{\gamma_n}(p_{\gamma_n}) \big]
\eeq
proved in appendix \ref{app:1DimChain}. Translational symmetry and formula \eqref{eq:Marg} lead finally to
\beqa
S(T \bar \phi) &=& \frac{N}{\rho} \Bigg\{  \sum_{s_{\Gamma_0}} \left[ \frac{1}{2^R} \sum_{{\Gamma^\prime} \subseteq \Gamma_0} \sum_{\mu=1}^M c_{\mu,\Gamma^\prime} \, \bar \phi_\mu \, \hat \phi_{\Gamma^\prime} (s_{\Gamma^\prime}) \right]  \log
\left[ \frac{1}{2^R} \sum_{{\Gamma^\prime} \subseteq \Gamma_0} \sum_{\mu=1}^M c_{\mu,\Gamma^\prime} \, \bar \phi_\mu \, \hat \phi_{\Gamma^\prime} (s_{\Gamma^\prime})\right]  \nonumber  \\
&-& \sum_{s_{\gamma_0}}\left[ \frac{1}{2^{R-\rho}} \sum_{\gamma^\prime\subseteq \gamma_0} \sum_{\mu=1}^M c_{\mu,\gamma^\prime} \, \bar \phi_\mu \, \hat \phi_{\gamma^\prime} (s_{\gamma^\prime})\right]  \log
\left[ \frac{1}{2^{R-\rho}} \sum_{\gamma^\prime\subseteq \gamma_0} \sum_{\mu=1}^M c_{\mu,\gamma^\prime} \, \bar \phi_\mu \, \hat \phi_{\gamma^\prime} (s_{\gamma^\prime})\right] \Bigg\} \; ,
\eeqa
where the $\hat \phi_\Gamma(s_\Gamma)$ are monomials used to expand the marginals as in \eqref{eq:Marg}, while $c_{\mu,\Gamma} = 1$ if it exists an $n $ such that $T_n \phi_\mu = \hat \phi_\Gamma$ and $c_{\mu,\Gamma} = 0$ otherwise. The couplings are given by
\beqa
g_\mu^\star &= & \sum_{s_{\Gamma_0}} \left[ \frac{1}{2^R} \sum_{{\Gamma^\prime} \subseteq \Gamma_0} c_{\mu,\Gamma^\prime}  \, \hat \phi_{\Gamma^\prime} (s_{\Gamma^\prime})\right]  \log
\left[ \frac{1}{2^R} \sum_{{\Gamma^\prime} \subseteq \Gamma_0} \sum_{\nu =1}^M c_{\nu,\Gamma^\prime} \, \bar \phi_\nu \, \hat \phi_{\Gamma^\prime} (s_{\Gamma^\prime})\right]  \nonumber  \\
&-& \sum_{s_{\gamma_0} }\left[ \frac{1}{2^{R-\rho}} \sum_{\gamma^\prime\subseteq \gamma_0} c_{\mu,\gamma^\prime} \, \hat \phi_{\gamma^\prime} (s_{\gamma^\prime})\right]  \log
\left[ \frac{1}{2^{R-\rho}} \sum_{\gamma^\prime\subseteq \gamma_0} \sum_{\nu=1}^M c_{\nu,\gamma^\prime} \, \bar \phi_\nu \, \hat \phi_{\gamma^\prime} (s_{\gamma^\prime})\right] \label{eq:Coupl1DChain}
\eeqa
while the inverse susceptibilities can be written as
\beqa
\chi^{-1}_{\mu,\nu} &=&   \frac{\rho}{N} \Bigg\{  \sum_{s_{\Gamma_0}} \frac{
\left[ \frac{1}{2^R} \sum_{{\Gamma^\prime} \subseteq \Gamma_0} c_{\mu,\Gamma^\prime}  \, \hat \phi_{\Gamma^\prime} (s_{\Gamma^\prime})\right] 
\left[ \frac{1}{2^R} \sum_{{\Gamma^\prime} \subseteq \Gamma_0} c_{\nu,\Gamma^\prime}  \, \hat \phi_{\Gamma^\prime} (s_{\Gamma^\prime})\right] 
}{
\bar p_{\Gamma_0} (T \bar \phi)
} \nonumber \\
&-&
\sum_{s_{\gamma_0}} \frac{
\left[ \frac{1}{2^R} \sum_{\gamma^\prime\subseteq \gamma_0} c_{\mu,\gamma^\prime}  \, \hat \phi_{\gamma^\prime} (s_{\gamma^\prime})\right] 
\left[ \frac{1}{2^R} \sum_{\gamma^\prime\subseteq \gamma_0} c_{\nu,\gamma^\prime}  \, \hat \phi_{\gamma^\prime} (s_{\gamma^\prime})\right] 
}{
\bar p_{\gamma_0} (T \bar \phi)
}\Bigg\} \; . \label{eq:Fisher1DChain}
\eeqa
The structure of this solution is analogous to the one of \eqref{eq:SolTree}, in the sense that once that the $S(p)$ is written as a sum of cluster entropies $S_{\Gamma_n}(p_{\Gamma_n})$ and $S_{\gamma_n}(p_{\gamma_n})$, the expression for the couplings is obtained by summing their separate contributions to the estimator $g^\star_\mu(T\phi)$. The main difference in this case is that one can exploit translational symmetry in order to equate the cluster contributions due to different unit cells. Thus, the number of sampled unit cells is $\sqrt{T N/\rho}$, rather that $\sqrt{T}$. This rules the convergence of the couplings to their asymptotic values: if the system is large translational symmetry allows to obtain reliable estimations of the couplings even with a small number of observations.

\section{Applications\label{sec:Appl}}
\begin{figure}[h] 
   \centering
   \includegraphics[width=4in]{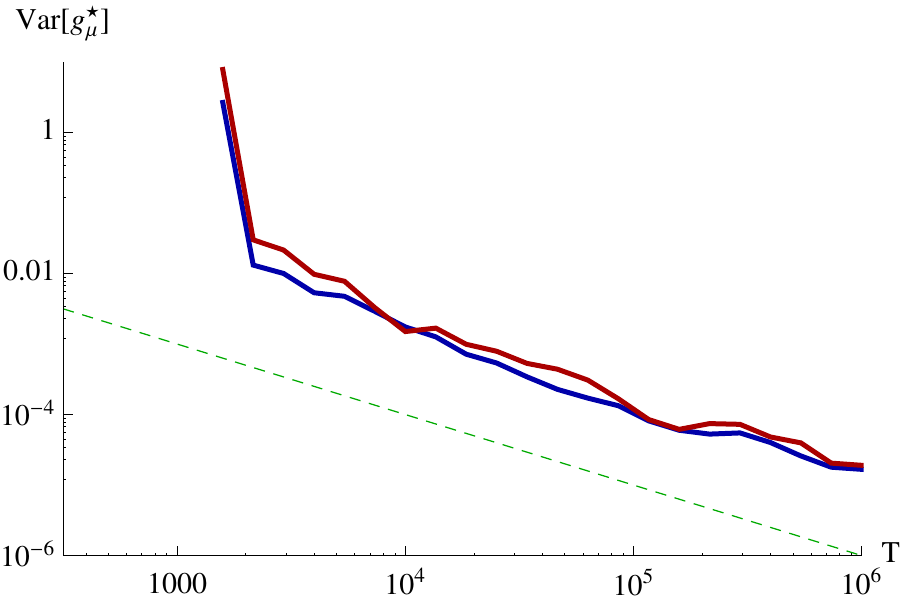} 
   \caption{Variance of the inferred couplings $h$ (red line) and $J$ (blue line) against the number of samples $T$ for a pairwise tree. The dashed green line plotted for reference indicates the expected scaling $1/T$.}
   \label{fig:TreeError}
\end{figure}

\begin{figure}[h] 
   \centering
   \begin{tabular}{cc}
   \includegraphics[width=3in]{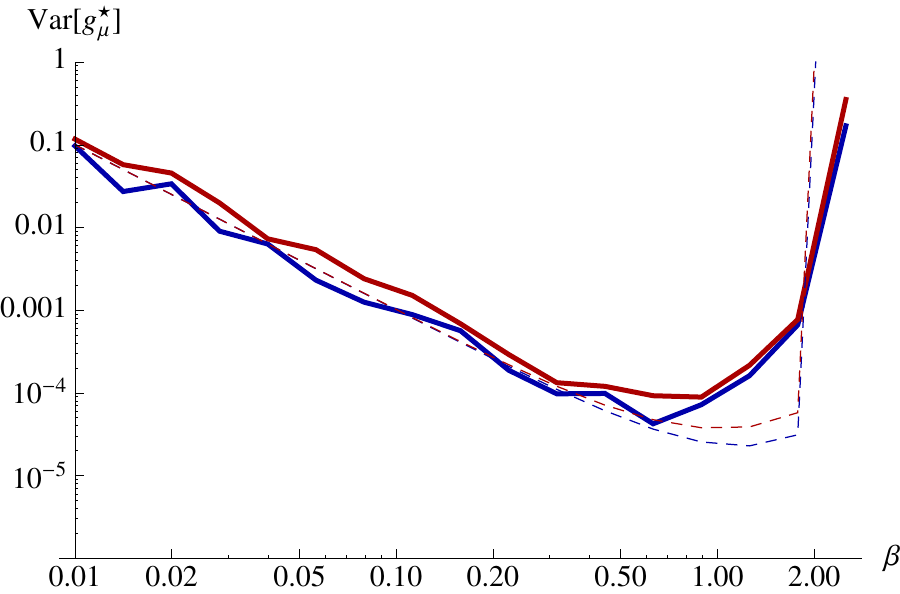}
   \includegraphics[width=3in]{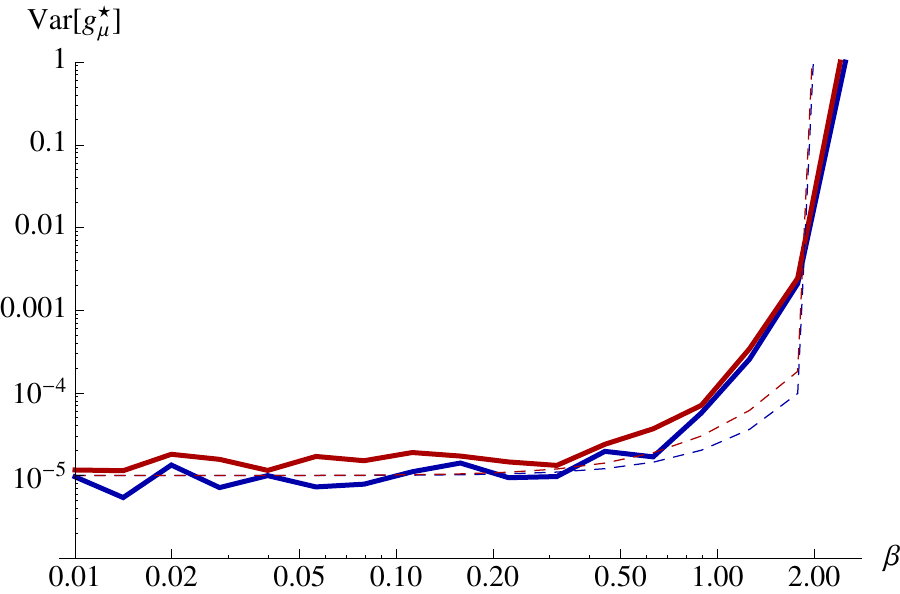} 
   \end{tabular}
   \caption{Variance of the inferred couplings $h^\star$ (red line) and $J^\star$ (blue line) against the inverse temperature $\beta$ for a pairwise tree, obtained by using $T=10^5$ MonteCarlo samples. Both the variances of $h^\star$ and $J^\star$ (left panel) and the ones of the products $h^\star \beta$ and $ J^\star \beta$ (right panel) are plotted. The dashed lines indicate the estimations of the error obtained by using expression \eqref{eq:FisherInfoPairTree}.}
   \label{fig:TreeVarBeta}
\end{figure}

\begin{figure}[h] 
   \centering
   \includegraphics[width=4in]{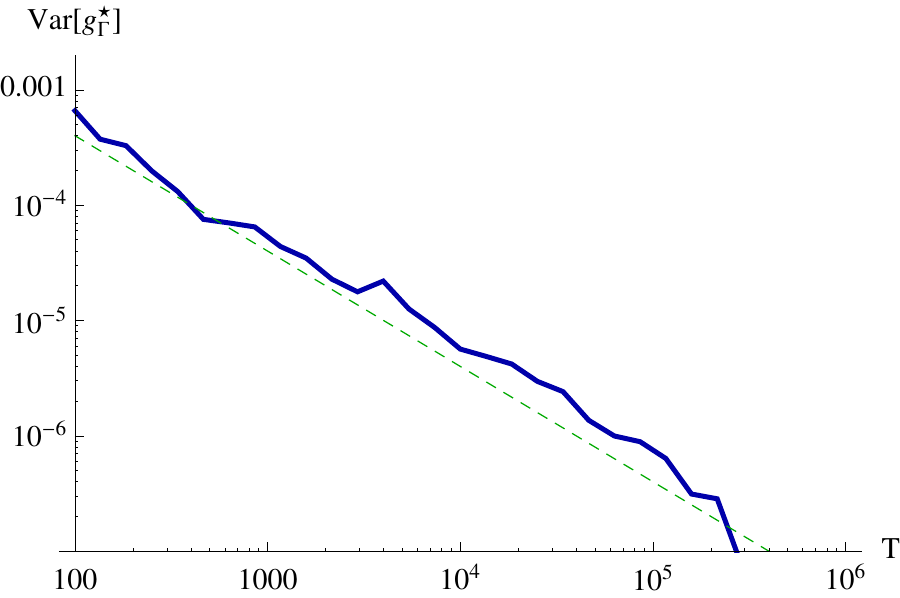}
   \caption{Variance of the inferred coupling vector $g^\star$ (blue line) plotted against the number of sampled unit cells $N T/\rho$, obtained by MonteCarlo sampling of a model describing a complete one-dimensional periodic chain of size $N=50$, range $R=4$ and periodicity $\rho=2$. The green dashed line shows the reference scaling $\rho/NT$.}
   \label{fig:1DimChainError}
\end{figure}

\begin{figure}[h] 
   \centering
   \begin{tabular}{cc}
   \includegraphics[width=3in]{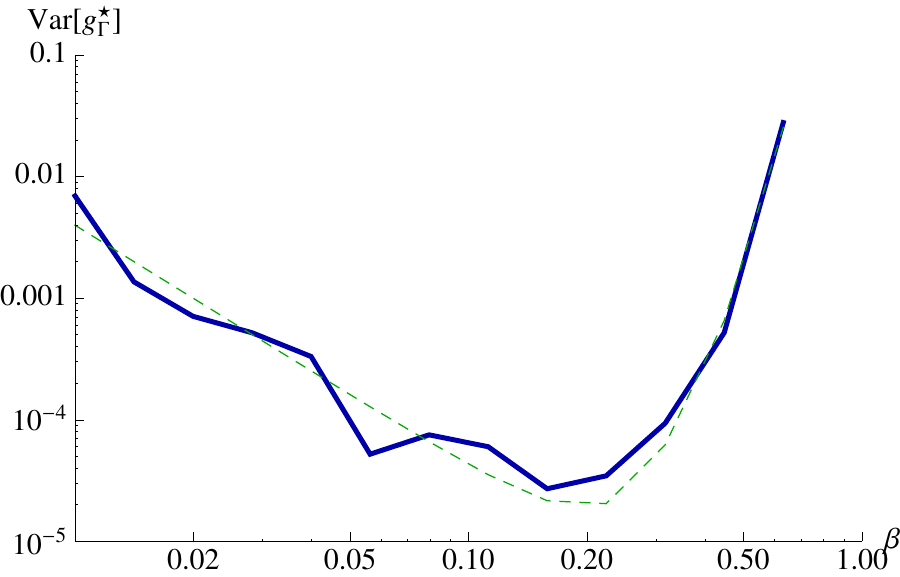}
   \includegraphics[width=3in]{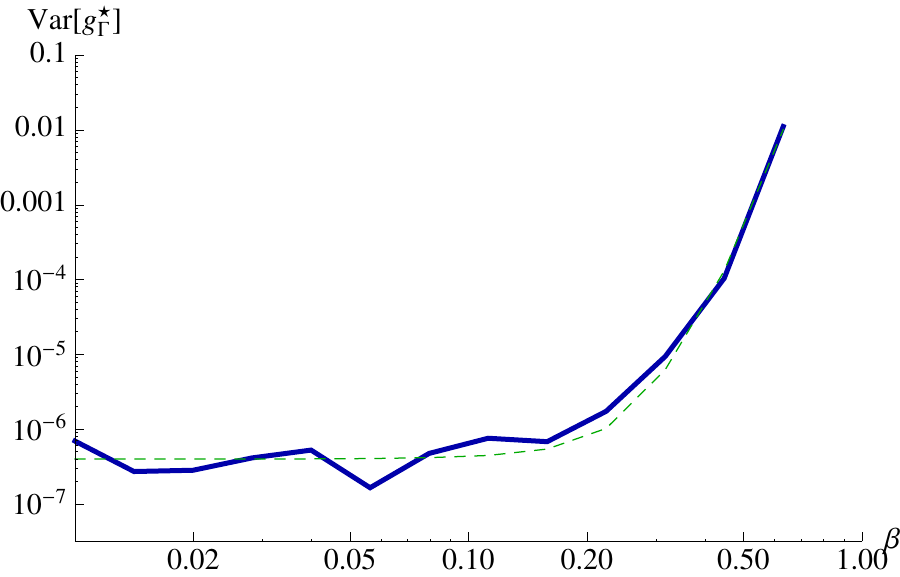} 
   \end{tabular}
   \caption{Variance of the inferred couplings $g^\star$ (blue line) against the inverse temperature $\beta$ for a one dimensional periodic chain. $T=10^5$ configurations have been sampled via MonteCarlo so to construct the empirical averages $\bar \phi$. Both the the results for the inferred couplings $g^\star$ (left panel), and the one for the products $\beta g^\star$ (right panel) are shown. The dashed line indicates the estimation of the error obtained by using equation \eqref{eq:Fisher1DChain}.}
   \label{fig:1DimChainVarBeta}
\end{figure}

In order to validate the results shown above, some examples involving synthetic datasets are discussed. In the case of a tree-like model, a set of $T=10^6$ configurations of a systems of size $N=50$ has been simulated. The configurations have been obtained by MonteCarlo sampling from a distribution $p(s)$ specified by a random tree-like graph $E$, while the couplings $J$ and $h$ have been uniformly drawn from the interval $[0,1]$. The variances of the inferred couplings for specific instances of the problem have been shown in figure \ref{fig:TreeError}, where they are compared against the expected scaling $1/T$. The plot indicates that formula \eqref{eq:SolTree} correctly predicts the value of the inferred couplings and of their fluctuations.
I have also considered the problem in which every coupling is multiplied by an \emph{inverse temperature} parameter $\beta$, in order to control the overall intensity of the fluctuations. The presence of strong noise corresponds to the regime of large $\beta$, while the one in which noise is suppressed is associated with the small $\beta$ region. In this example the couplings $J$ and $h$ have been uniformly drawn in $[0,1]$ and multiplied by a $\beta$ parameter in the range $[0.01,10]$. Subsequently, $T=10^5$ samples have been extracted via MonteCarlo.  In figure \ref{fig:TreeVarBeta} the variance of the inferred couplings is plotted against the inverse temperature $\beta$.
One can notice from the plots that, as the inverse problem depends just upon the products $\beta J$ and $\beta h$, it is not possible to distinguish between an overall strength of the fluctuations a the temperature parameter $\beta$. Hence, while the lowest inference error for the products $\beta J$ and $\beta h$ is obtained for $h_i=J_{ij} = 0$, the lowest error for the values $J$ and $h$ (given a fixed, known value for $\beta$) is obtained by finding a compromise between maximum signal (favoring high couplings) and minimum noise (favoring high temperature, or equivalently low $\beta$). One also finds that the quality of the reconstruction degrades progressively in the large $\beta$ regime by increasing the inverse temperature, until eventually some of the two-state marginals are not sufficiently sampled, in which case the inferred couplings diverge. 
The same type of study has been carried on for the model \eqref{eq:1DimChain}. As an example, I show the results obtained for a system of size $N=50$, characterized by interactions of range $R=4$ and periodicity parameter $\rho=2$. A number of configurations $T=10^6$ have been simulated via MonteCarlo for a system with random interactions ($g_\mu$ uniformly extracted in $[0,0.01]$). The results are shown in figure \ref{fig:1DimChainError}, where the variance of the inferred couplings is plotted against the expected scaling $\rho/NT$. Also for the one dimensional case I analyzed the effect of an inverse temperature parameter $\beta \in [0.01,1]$ modulating an interaction strength $g_\mu$ randomly drawn from the interval $[0,1]$. Figure \ref{fig:1DimChainVarBeta} shows the errors in the reconstructed couplings as a function of the $\beta$ parameter.
Notice that the error on $\beta g^\star$ is minimum for a flat distribution ($\beta = 0$), while the one on $g^\star$ is minimized by an intermediate value of $\beta$. The quality of the reconstruction is determined by how uniformly is the $R$-spin marginal sampled. If local fluctuations are not sampled well-enough, the error on the inferred couplings is large as predicted by equation (\ref{eq:Fisher1DChain}). As observed above, it is not necessary to probe global fluctuations of the system in order to accurately reconstruct the couplings.

\section{General structure of exact solutions\label{sec:General}}
What has been shown for models \eqref{eq:IsingMod} and \eqref{eq:1DimChain} can be easily extended to more general scenarios. In particular any factorization property for the probability distribution $p(s)$ analogous to \eqref{eq:TreeFact} and \eqref{eq:ChainFact} breaks the entropy in a set of cluster contributions of the form
\beq
S(p) = \sum_{\Gamma \in \mathcal G} c_\Gamma S_\Gamma(p_\Gamma) \, ,
\eeq
where $\mathcal G$ is a suitably chosen collection of clusters associated with a set of $\{c_\Gamma\}_{\Gamma \in \mathcal G}$ coefficients.
Equation \eqref{eq:Marg} implies that cluster entropies $S_\Gamma(p_\Gamma)$ are local in the empirical observables associated with $\Gamma$.
Hence, if the model is large enough (i.e., if $\phi$ includes all the monomials located within the clusters $\Gamma \in \mathcal G$),
the cluster entropies can be expressed as functions of the observables $\bar \phi$ conjugated to the couplings $g$.
In this case the following facts hold:
\begin{itemize}
\item{
Due to linearity of the derivative any inferred coupling $g^\star_\mu(\bar \phi)$ receives contributions just by clusters containing its conjugated potential $\phi_\mu(s)$.
}
\item{The matrix $\chi^{-1}$ is sparse, so that by shifting any of the empirical observables $\bar \phi_\mu$ the only clusters which are affected are the ones including variables contained inside the monomial $\phi_\mu(s)$.
}
\end{itemize}
The expressions \eqref{eq:FisherInfoPairTree} and \eqref{eq:Fisher1DChain} for the stability matrix $\chi^{-1}$ have then a clear interpretation in term of cluster contributions: if the clusters necessary to describe $p(s)$ are  uniformly sampled (i.e., the empirical frequencies $\bar p_\Gamma$ are approximately flat), the error on the inferred couplings is small. Conversely, as soon as one ore more configurations $s_\Gamma$ are not sampled, the asymptotic error diverges. This is consistent with the observation that if a $p_\Gamma(s_\Gamma)$ assigns zero mass to some configurations, infinite couplings are required to describe the marginals (in that case either some regularization can be used to enforce finiteness of the solution, or infinite solutions may be accepted as describing a hard constraint on the set of accessible configurations).
As a further comment, notice that a reliable estimation of the couplings $g$ doesn't require sampling global fluctuations in the configuration space: it is simply necessary to sample fluctuations which are local with respect to clusters in $\mathcal G$ in order to accurately estimate the couplings $g$. \\ \\
These ideas can be useful even in the case in which the entropy doesn't strictly break in the sum of small cluster terms. In particular it has been observed in \cite{Cocco:2011vn,Cocco:2012uq} that for several models it is possible to account for a large fraction of the entropy $S(p)$ by using just a few local terms (corresponding to the remark that in many cases the matrix $\chi^{-1}$ is approximately sparse). In this case fast, although approximate, inference schemes could be obtained by using these techniques. \\ \\
Notice that the price to pay in order to analytically solve this inverse problem is the introduction of a potentially high number of parameters: a coefficient has to be fitted for each of the monomials included in any of the clusters $\Gamma \in \mathcal G$. In particular, the number of parameters to be considered is bound by $M \leq |\mathcal{G}| \times  2^{\max_{\Gamma \in \mathcal G} |\Gamma|}$. Thus, even if a number of clusters polynomial in $N$ is usually sufficient to properly describe the entropy of a system, their size enters exponentially in the number of fitted parameters.

\section{Conclusions\label{sec:Concl}}
The task of determining a probability distribution whose associated momenta match a given set of values is a known hard problem, with relevant implications in the field of high-dimensional inference. In this work I have shown that its exact solution can be found whenever the probability density for the system has a factorized structure. The sparsity of the Fisher information matrix $\chi^{-1}(\bar \phi)$ and the additivity of the entropy can be shown to rely upon this independence property. These result is completely general (in particular, it is independent of the interaction structure of the system). In order to provide illustrative examples, these ideas have been used to solve two specific inverse problems: the inverse Ising model on a tree and the one-dimensional periodic chain with arbitrary order interactions. The interactions of both systems can be accurately reconstructed if the clusters describing the system are uniformly sampled, while the inferred couplings diverge as soon as some relevant micro states are unobserved.
These findings have been confirmed by numerical simulations, which show an excellent agreement with the analytical predictions. Techniques for the approximate solution of generic inverse problems could in principle be derived by developing the methods presented in this work.


\appendix
\section{Factorization property for the one-dimensional chain\label{app:1DimChain}}
\begin{figure}[h] 
   \centering
   \includegraphics[width=4in]{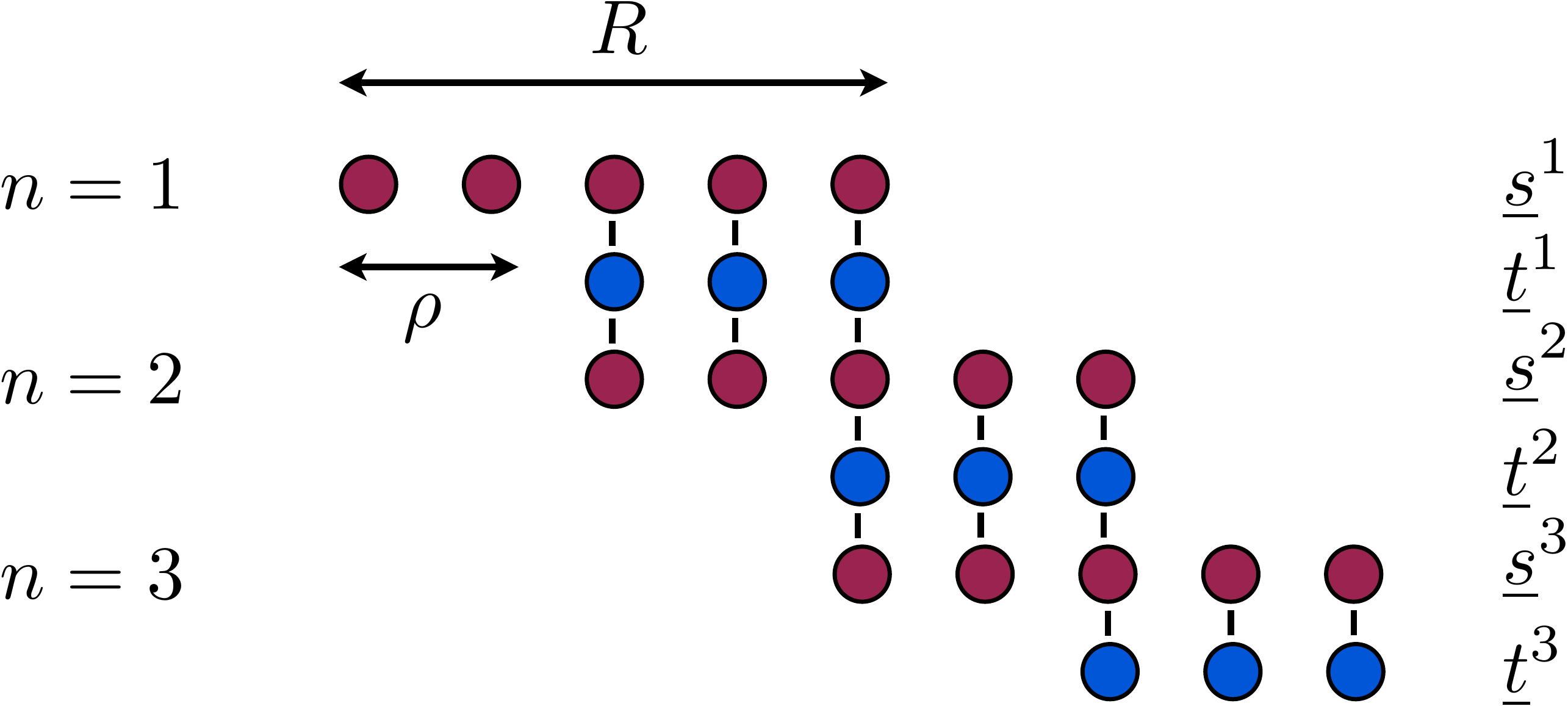} 
   \caption{Two dimensional auxiliary model $p^\lambda(s)$ associated with the original distribution $p(s)$ describing a one-dimensional periodic chain.}
   \label{fig:2DMap}
\end{figure}

I show in the following that for a one-dimensional periodic chain defined as in \eqref{eq:1DimChain}, the factorization property
\beq
 p(s) = \prod_{n=0}^{N/\rho-1} \frac{p_{\Gamma_n}(s^{\Gamma_n})}{p_{\gamma_n}(s^{\gamma_n})} \; 
\eeq
holds, where $\Gamma_n = \{ n\rho+1, \dots,n\rho+R\}$ and $\gamma_n=\{ (n+1)\rho+1,\dots, n\rho+R\}$.
To obtain this result, one can consider the auxiliary two-dimensional model defined by the log-probability
\beqa
\log p^\lambda (s,t) &=& - \log Z(g) + \sum_{n=0}^{N/\rho-1} \sum_{\mu \in \phi}  g_\mu \phi_\mu(s_{1+n\rho}^n,\dots,s_{R+n\rho}^n) \nonumber \\
&+& \lambda \sum_{n=0}^{N/\rho-1} \sum_{i=(n+1)  \rho + 1}^{n \rho + R} \left[ (t_i^n - s_i^n)^2+ (t_i^n - s_i^{n+1})^2 \right] \; ,
\eeqa
in which the configuration space contains the original degrees of freedom are $s_i^n \in \{ -1,1\}$ (with $n=0,\dots,N/\rho-1$ and $i=1+n\rho,\dots, R+n\rho$) and the auxiliary ones $t_i^n \in \{ -1,1\}$ (with $n=0,\dots,N/\rho-1$ and $i=1+(n+1)\rho,\dots, R+n\rho$). The relation between the original model and the auxiliary one is sketched in figure \ref{fig:2DMap}. In particular the coupling $\lambda$ controls the strength of the bond in the auxiliary dimension (labeled by $n$), so that the limit $\lambda \to \infty$ describes the original chain with the obvious identification $s^n_i \to s_i$ and $t^n_i \to s_i$.
By defining the row variables $\underline s^n = \{ s_i^n\}_{i=1+n\rho}^{i=R+n\rho}$ and $\underline t^n = \{ t_i^n\}_{i=1+(n+1)\rho}^{i=R+n\rho}$, the log-probability for the two dimensional model can be written as
\beq
\log p^\lambda (s,t) = -\log Z^\lambda(g) - \sum_{n=0}^{N/\rho-1} \bigg[ \mathcal H^\lambda_n(\underline s^n)  + \mathcal H_{n,n}^\lambda(\underline s^n, \underline t^n) + \mathcal H_{n,n+1}^\lambda (\underline t^n, \underline s^{n+1})\bigg]   \; . \label{eq:2DMapTree}
\eeq
Hence the distribution over the degrees of freedom $\underline s^n$ and $\underline t^n$ defines a \emph{tree}, because only successive row of variables interact\footnote{Periodic boundary conditions enforce the presence of a single loop of length $N$, so that the model is not exactly a tree. Nevertheless, for $N$ large enough and for $g$ sufficiently distant from critical points of the model, if any, the presence of such loop can be neglected.}. The marginals associated with the clusters $\Gamma_n$ and $\gamma_n$ can be used in order to express the probability $p^\lambda(s,t)$ as
\beq
p^\lambda(s,t) = \frac{
\prod_n p^\lambda_{\Gamma_n \cup  \gamma_n} (\underline s^{n} ,\underline t^{n} ) p^\lambda_{\gamma_n \cup \Gamma_{n+1}} (\underline t^n ,\underline s^{n+1} )
}{
\prod_n p^\lambda_{\Gamma_n} (\underline s^n) p^\lambda_{\gamma_n} (\underline t^n)
} \; ,
\eeq
where for the two-dimensional model $\Gamma_n$ and $\gamma_n$ are analogously defined. By taking the $\lambda \to \infty$ limit, the identification
\beqa
p^\lambda_{\Gamma_n \cup  \gamma_n} (\underline s^{n} ,\underline t^{n} )&\xrightarrow[\lambda \to \infty]{}& p_{\Gamma_n}(s_{n\rho +1} ,\dots , s_{n \rho +R}) \\
p^\lambda_{\Gamma_n} (\underline s^n) &\xrightarrow[\lambda \to \infty]{}&  p_{\Gamma_n}(s_{n\rho +1} ,\dots , s_{n \rho +R}) \\
p^\lambda_{\gamma_n} (\underline t^n) &\xrightarrow[\lambda \to \infty]{}&  p_{\gamma_n} (s_{(n+1)\rho +1} ,\dots , s_{n \rho +R}) \; .
\eeqa
allows to recover the factorization property \eqref{eq:ChainFact}.


\bibliographystyle{unsrt}

\bibliography{ExactInference}

\end{document}